\documentclass{amsproc}

\newtheorem{theorem}{Theorem}[section]

\newtheorem{proposition}[theorem]{Proposition}
\theoremstyle{definition}

\theoremstyle{remark}
\newtheorem{remark}[theorem]{Remark}
\newtheorem{remarks}[theorem]{Remarks}
\numberwithin{equation}{section}

\begin{document}

\title{Recent Results on Non--Adiabatic
Transitions in Quantum Mechanics}

\author{George A. Hagedorn}
\address{Department of Mathematics and
Center for Statistical Mechanics and Mathematical Physics,
Virginia Tech,
Blacksburg, Virginia, 24060--0123,
USA}
\email{hagedorn@math.vt.edu}
\thanks{Supported in part by NSF Grant DMS--0303586.}
\author{Alain Joye}
\address{Institut Fourier,
Unit\'e Mixte de Recherche CNRS--UJF 5582,
Universit\'e de Grenoble I,
BP 74,
F--38402 Saint Martin d'H\`eres Cedex,
France}
\email{alain.joye@ujf-grenoble.fr}

\date{31 August 2005.}


\begin{abstract}
We review mathematical results concerning exponentially small
corrections to adiabatic approximations and Born--Oppenheimer
approximations.
\end{abstract}

\maketitle


\section*{Introduction}
The goal of this paper is to review 
recent results on exponentially small 
non-adiabatic transitions in quantum mechanics.
In Section 1, we provide background information about
adiabatic approximations. In Section 2, we discuss the determination
of non--adiabatic scattering transition amplitudes. In Section 3,
we describe the time development of exponentially
small non--adiabatic transitions.
In Section 4, we turn to exponentially accurate Born--Oppenheimer
approximations. Then in Section 5, we discuss the
determination of non--adiabatic corrections to Born--Oppenheimer
approximations in a scattering situation.


\section{Adiabatic Background}

The adiabatic approximation in quantum mechanics concerns
the time--de\-pen\-dent Schr\"odinger equation when the Hamiltonian
depends on time, but varies on a very long time scale.
Mathematically, this situation corresponds to the singularly
perturbed initial value problem
\begin{equation}\label{sch}
i\,\epsilon\,\partial_t\psi_\epsilon(t)\ =\ H(t)\,\psi_\epsilon(t),
\qquad \psi_\epsilon(0)\ =\ \phi(0),
\end{equation}
where $t\in \mathbb R$, $\epsilon$ is a small parameter, and
$\psi_\epsilon(t)$ belongs to a separable Hilbert space 
${\mathcal H}$.

In a simple situation, the adiabatic approximation relies
on two basic assumptions.
The first is a regularity condition on the Hamiltonian:\\

\noindent {\bf R:}\quad {\em The Hamiltonian $H(t)$
is a bounded self-adjoint operator on {$\mathcal H$} that
depends smoothly on $t$.}\\

We denote the unitary propagator corresponding
to (\ref{sch}) by $U_\epsilon(t,s)$. It satisfies
\begin{equation}\label{schu}
i\,\epsilon\,\partial_t U_\epsilon(t,s)\ =\ 
H(t)\,U_\epsilon(t,s),\qquad  U_\epsilon(s,s)\mathbb\ =\ I
\end{equation}
for all $(t,s)\in\mathbb R^2$

The second assumption is that there is a gap in the spectrum
of $H(t)$ at all times $t$: \\

\noindent {\bf G:}\quad {\em The spectrum of $H(t)$, denoted by
$\sigma(t)\subset \mathbb R$, can be decomposed as
$$\sigma(t)\ =\ \sigma_1(t)\cup\sigma_2(t),\quad  \mbox{with}\quad
\inf_{t\in \mathbb R}\,dist (\sigma_1(t),\,\sigma_2(t))\ =\ g>0.$$
}

Under this hypothesis there exists a self-adjoint spectral projection
$P(t)$ corresponding to $\sigma_1(t)$, which depends smoothly on $t$.

In this situation, the Adiabatic Theorem of Quantum Mechanics says
the following:
\begin{theorem}
Let $H(t)$ satisfy {\bf R} and {\bf G}, and let $P(t)$ be as above.
Then, for small $\epsilon$,
\begin{equation}
\|\,(\mathbb I -P(t))\,U_\epsilon(t,s)\,P(s)\,\|\ =\ O(\epsilon),
\end{equation}
for any $(t,s)$,
where the error term is sharp.
\end{theorem}
In other words, for small $\epsilon$, the quantum evolution follows
the isolated spectral subspaces of the Hamiltonian, up
to an error of order $\epsilon$. We call the error term the
transition amplitude between the subspaces $P(s)\mathcal H$ and
$(\mathbb I-P(t))\mathcal H$.

One modern proof actually yields a stronger result. Let us sketch it
for later reference. 
Let $V(t,s)$ be the ($\epsilon$-dependent)
unitary evolution defined by
\begin{equation}\label{v}
i\,\epsilon\,\partial_t V(t,s)\ =\
\left( H(t)-i\epsilon [\partial_tP(t),\,P(t)]\right)
\,V(t,s)\quad\mbox{and}\quad
V(s,s)\ =\ \mathbb I.
\end{equation}
It is not difficult to verify the intertwining property
\begin{equation}\label{inter}
V(t,s)\,P(s)\ =\ P(t)\,V(t,s),
\qquad\mbox{for all}\ (t,s)\in\mathbb R^2.
\end{equation}
A generalized integration by parts procedure then establishes
the estimate $\|U_\epsilon(t,s)-V(t,s)\|=O(\epsilon |t-s|)$, which 
implies the result.

Historically, the
Adiabatic Theorem goes back to Born and Fock
\cite{bf} who proved it in 1928 for matrix valued Hamiltonians
with non-degenerate eigenvalues.
In the fifties, Kato \cite{k} proved the statement for general
Hamiltonians with an isolated eigenvalue $E(t)$.
Thirty years later, Nenciu showed the isolated eigenvalue $E(t)$
could be replaced by an isolated piece of spectrum $\sigma_1(t)$.
The extension to unbounded operators
was proven by Avron, Seiler and Yaffe in \cite{asy}.

Although we shall always assume the gap hypothesis {\bf G},
there are recent versions of the
adiabatic theorem that hold without it.
Instead, these results require the existence of
a sufficiently regular, finite rank family of projectors $P(t)$
that satisfy $P(t)H(t)=E(t)P(t)$ for all $t$.
Here $E(t)$ is an eigenvalue of $H(t)$, and $P(t)$ is a spectral
projector of $H(t)$ for almost all $t$. See {\em e.g.},
\cite{bornemann}, \cite{avron-elgart}, \cite{teufelgap}.
In this situation, 
the transition amplitude can go to zero arbitrarily slowly
as $\epsilon$ tends to zero.

Since the early thirties, it has been known that
the adiabatic theorem could be substantially
improved under certain circumstances.
Zener \cite{zener} considered (\ref{sch}) for the $2\times 2$
matrix Hamiltonian
\begin{equation}
H(t)\ =\ \frac12\ \left(\,
\begin{matrix}t &\delta \\ \delta &-t\end{matrix}\,
\right),
\end{equation}
whose spectrum is
$\sigma(t)\ =\ \{\,\pm\, \frac12\, \sqrt{t^2+\delta^2}\,\}$.
For this Hamiltonian, (\ref{sch}) can be solved exactly in terms
of parabolic cylinder functions, for any $\delta$ and 
any $\epsilon>0$.
In particular, in the scattering regime where the initial
and final times $s$ and $t$ tend to $-\infty$ and $+\infty$ respectively,
the transition amplitude takes the simple form
\begin{equation}\label{lz}
\mathcal A(\epsilon)\ :=\
\lim_{s\rightarrow -\infty
\atop 
t\rightarrow +\infty}
\|\,(\mathbb I -P(t))\,U_\epsilon(t,s)\,P(s)\,\|\ =\ 
e^{-\pi\delta^2/(4\epsilon)}.
\end{equation}
Zener's analysis shows that the transition
amplitude decreases from its order $\epsilon$ value for finite
times $(s,t)$ to an exponentially small value in $1/\epsilon$ in the
scattering limit.
Shortly after Zener's result, Landau \cite{lan} argued that (\ref{lz})
should also hold for more general analytic $2\times 2$ real symmetric
Hamiltonians whose non-degenerate eigenvalues $E_1(t)$ and $E_2(t)$
displayed an avoided crossing with gap $\delta$, {\em i.e.}, when
\begin{equation}\label{avcr}
E_2(t)-E_1(t)\ \simeq \,\sqrt{(t-t_0)^2+\delta^2}.
\end{equation}
Here $\delta$ is a small, but fixed, extra parameter in the problem.
These considerations gave rise to the famous Landau--Zener formula,
which showed that non-adiabatic transitions between spectrally
isolated subspaces belonged to the realm of exponential asymptotics.

This Landau--Zener mechanism is the main concern of this review.


\section{Exponentially Small Transitions in the Adiabatic
Approximation}

From a mathematical point of view, the first exponential bounds on
transition amplitudes in a general framework were obtained in 
\cite{jp0}.
If we assume for simplicity that the Hamiltonian is bounded,
this result requires the following
regularity and scattering hypotheses:\\

\noindent {\bf A:}\quad {\em There exists $\mu>0$
such that the map $t\mapsto H(t)\in {\mathcal L(\mathcal H)}$
has an analytic extension to the strip
$S_\mu=\{ z\in\mathbb C \,:\ |\mbox{Im}(z)|\leq \mu\}$. 
}\\

\noindent {\bf S:}\quad {\em There exists $\nu >1$,
two bounded self-adjoint operators  $H(\pm\infty)$,
and a constant $c$, such that as $t\rightarrow
\pm\infty$,
$$
\sup_{|s|\leq \mu}\,\|H(t+is)-H(\pm\infty)\|\
\leq\ \frac{c}{\langle t\rangle^{\nu}},
$$
where $\langle t\rangle =\sqrt{1+t^2}$.
}

\begin{theorem}\label{exes}\cite{jp0}
Assume $H(t)$ satisfies  {\bf A},  {\bf S} and {\bf G}. Then,
there exist $C$ and $\Gamma >0$ such that
$\mathcal A(\epsilon)\ \leq\ C 
e^{-\Gamma/\epsilon}$, for small $\epsilon>0$.
\end{theorem}
The original proof in \cite{jp0} establishes properties of
solutions to the Schr\"odinger equation for complex $t$,
by using complex WKB techniques. 
Similar results were subsequently proven using other 
methods. See {\em e.g.}, \cite{nenciu2}, \cite{sj}, \cite{jp1}, 
and \cite{martinez}.

A rigorous asymptotic formula for the transition amplitude across
the gap of a two-level system was proven in \cite{jkp}.
In addition to the assumptions above, the proof 
required supplementary hypotheses on the behavior in the complex plane 
of the so-called Stokes lines of the problem.

This hypothesis, that we call {\bf SL}, is complicated, and we
shall not attempt to describe it here.
It is typical of the 
complex WKB method and is discussed in detail in \cite{jkp}.
\begin{theorem}
Let
$$
H(t)\ =\ \frac 12 
\left(\,
\begin{array}{cc}Z(t)&X(t)+i\,Y(t)\\ X(t)-i\,Y(t)&-Z(t)\end{array}
\,\right),
$$
be a $2\times 2$ hermitian matrix that satisfies assumptions
{\bf A}, {\bf S} and {\bf G}.
Further assume that the analytic extension
of $\rho(t)=\sqrt{X^2(t)+Y^2(t)+Z^2(t)}$ has only one
conjugate pair of zeros, $\{z_0,\,\bar{z}_0\}$, in $S_\mu$,
and that the supplementary hypothesis
{\bf SL} is satisfied. Then, there exist $G\neq 0$
and $\gamma>0$, such that, as $\epsilon\rightarrow 0$,
\begin{equation}\label{prex}
\mathcal A(\epsilon)\ =\ G\,e^{-\gamma/\epsilon}\,(1+O(\epsilon)),
\end{equation}
where 
$\gamma=\left|\mbox{Im}\, \int_\zeta\rho(z)\, dz\right|/2$,
and $\zeta$ is a loop based at the origin which encircles $z_0$.
\end{theorem}
\begin{remarks}
1) The point $z_0$ is a complex crossing point for the analytic 
continuations of the eigenvalues of $H(t)$, and the decay rate 
$\gamma$ is the one predicted by Landau.\\
2) The prefactor $G$ is equal to one for generic
real-valued symmetric Hamiltonians.
It can take other values for generic hermitian
matrices, as was independently discovered by Berry \cite{be1}.\\
3) Several variants and improvements of this theorem have been
proven. See \cite{J1}, \cite{jp1}, and \cite{AJ1}.
Extensions to more general scattering systems were proved in 
\cite{AJP}, \cite{mn}, and \cite{AJ2}. In particular, 
\cite{AJ1, AJ2} show that in an avoided crossing regime, 
hypothesis {\bf SL} is automatically satisfied. See also \cite{iamp}.
\end{remarks}
\vspace{2mm}

These results were first proven by means of the complex WKB
technique. Another popular and fruitful method in exponential
asymptotics is the so-called optimal truncation method. In this
method, one first derives an asymptotic expansion of the desired 
quantity in powers of $\epsilon$.
Next, one carefully estimates the difference
between the exact quantity and the truncation of the expansion
after $n$ terms. This error estimate depends on both 
$n$ and $\epsilon$, and if it is bounded by
$c_0\,c_1^n\,n!\,\epsilon^n$,
one can choose $n$ to depend on $\epsilon$
in an optimal way to obtain an error estimate that is
$O(e^{-\Gamma(g)/\epsilon})$. The optimal $n$ satisfies
$n\simeq g/\epsilon$, where $g>0$ is sufficiently small. 

Nenciu \cite{nenciu2} first implemented these ideas in a general
adiabatic context. His work followed contributions by Garrido \cite{g}
and Sancho \cite{s} in the sixties.
The optimal truncation technique was also used in \cite{jp1}, and an 
elementary exposition of it is provided in \cite{hagjoy7}. 
More recently, \cite{NenSor} and \cite{teufelbook}
further adapted and generalized these ideas
to deal with space-time adiabatic theory.
 
Let us briefly present this circle of ideas through
the construction used in \cite{jp1}.
For a Hamiltonian $H(t)$ satisfying {\bf R} and {\bf G},
we set $H_0(t)=H(t)$, $P_0(t)=P(t)$ and
$K_0(t)=[\partial_tP_0(t),P_0(t)]$.
Then, we iteratively construct a sequence of
self-adjoint operators as follows.
For any integer $q>0$ and $\epsilon$ sufficiently small, we define
\begin{eqnarray}\label{iter}
H_q(t,\epsilon)&=&H(t)-\,i\,\epsilon\,K_{q-1}(t,\epsilon),\qquad
\mbox{ and }\\\nonumber
K_q(t,\epsilon)&=&[\partial_tP_q(t,\epsilon),\,P_q(t,\epsilon)],
\end{eqnarray}
where $P_q(t,\epsilon)$ is the spectral projector of $H_q(t,\epsilon)$ 
corresponding to $P_0$ in the limit $\epsilon\rightarrow 0$. 
Perturbation techniques ensure that this
scheme is well-defined for any $q$, provided $\epsilon$ is small.
Furthermore, hypothesis {\bf S} implies 
\begin{eqnarray}\label{sss}
P_q(t,\epsilon)&=&P_0(t)\ +\ O(\epsilon),\quad
\mbox{as }\ \epsilon\rightarrow 0,
\qquad\mbox{and}\\ \nonumber
\lim_{t\rightarrow \pm\infty}P_q(t,\epsilon)&=&P(\pm\infty),
\quad \mbox{for any fixed } \epsilon . 
\end{eqnarray}
We next introduce the evolution $V_q(t,s)$
associated with $H_q(t,\epsilon)$
as the solution to (\ref{v}), with $H_q$ in place of $H$ and 
$K_q$ in place of $K_0$.
By construction, the intertwining relation 
(\ref{inter}) holds with $V_q$ and $P_q$ in place of $V$ and $P$.
The motivation for this construction is that it yields the
estimate
\begin{equation*}
\|U_\epsilon(t,s)-V_q(t,s)\|\ \leq\
\int_s^t\,\|K_q(u,\epsilon)-K_{q-1}(u,\epsilon)\|\, du
\ \leq\  |t-s|\,\beta_q\,\epsilon^{q+1},
\end{equation*}
for small $\epsilon$,
where $\beta_q$ is finite. When $H(t)$ satisfies the 
analyticity hypothesis {\bf A}, one can further control $\beta_q$ as a 
function of $q$ and prove the bound 
$\beta_q\leq c_0\,c_1^q\,q!$.
The optimization over $q$ can then be performed. 
This implies the following theorem.

\begin{theorem} \cite{nenciu2,jp1}
Assume $H(t)$ satisfies {\bf A} and {\bf G}. 
There exist $\Gamma>0$ and an orthogonal projector 
$P_*(t,\epsilon)=P(t)+O(\epsilon)$
such that 
$$
\|(\mathbb I -P_*(t,\epsilon))U_\epsilon(t,s)P_*(s,\epsilon)\|
\ =\ O(e^{-\Gamma/\epsilon}\,|t-s|).
$$
\end{theorem}
\begin{remarks}
1. This result shows that although the transitions between instantaneous 
eigenprojectors of $H(t)$ are of order $\epsilon$ for finite times, there
exists another set of orthogonal projectors which are better suited to 
the adiabatic approximation, in the sense that transitions between them
are always exponentially small. \\
2. We call the projectors $P_*(t,\epsilon)$ optimal adiabatic projectors.
They are not uniquely defined. \\
3. We recover Theorem \ref{exes} if we further
assume {\bf S}. See (\ref{sss}).\\
4. The construction above was used in \cite{jp1} and \cite{AJ1} to reduce 
the study of transitions between two isolated levels in the spectrum
of a general Hamiltonian to an effective reduced $2$-level problem.
\end{remarks}

%


\section{Time Development of Exponentially Small
Non--Adiabatic Transitions}

Berry \cite{be1} pushes these ideas further for two-level systems. 
For such systems, a one-dimensional projector corresponds to an
orthonormal  basis of $\mathbb C^2$, with one vector in the range
and one vector in the kernel of the projector. 
Berry proposes studying the time development of the transition 
amplitude in this time-dependent basis. His heuristic 
arguments \cite{be1} are supported by spectacular numerical 
results \cite{lb}.
These  show that the transition amplitude passes from its
initial value of zero to its final exponentially small value
(\ref{prex}) in a universal way described by an error function. 

The first rigorous results \cite{hagjoy8} on this question
were obtained for the particular $2\times 2$
Hamiltonian function
\begin{equation}\label{H} H(t)\ =\
\frac 1{2\,\sqrt{t^2+\delta^2}}\
\left(\,\begin{array}{cc}\delta&t\\t&-\delta\end{array}\,\right),
\end{equation}
for fixed $\delta>0$.
The eigenvalues are $E_{1}(t)=1/2$ and $E_2(t)=-1/2$
for every $t$.

This Hamiltonian can be viewed as the Landau--Zener
Hamiltonian modified to keep its eigenvalues
constant and with the gap normalized to the value 1.

The notion of avoided crossing has been replaced by the
singularities of the Hamiltonian at $t\,=\,\pm\,i\,\delta$.
These points govern the transitions between the two levels
and yield the transition amplitude 
$\sqrt2\,e^{-\delta/\epsilon}$. See \cite{J1}. The time development
is described by the following result.

\begin{theorem}\label{supertheorem}\cite{hagjoy8}
Let $H(t)$ be given by (\ref{H}) and
let $\varphi_1(t)$ and $\varphi_2(t)$ be smooth real-valued
normalized eigenvectors corresponding to the eigenvalues 
$1/2$ and $-1/2$, respectively.
Then for any $\mu\in(0,1/2)$,

\noindent
{\bf 1.} There exist vectors
$\chi_1(\epsilon,\,t)$ and  $\chi_2(\epsilon,\,t)$ that satisfy the
Schr\"odinger equation (\ref{sch}) up to errors of order
$e^{-\delta/\epsilon}$
and correspond to the eigenstates $\varphi_1(t)$ and $\varphi_2(t)$ in
the sense that
\begin{equation}
\lim_{|t|\rightarrow \infty}\ 
|\,\langle\,\varphi_j(t),\,\chi_j(\epsilon,\,t)\, \rangle\,|\ =\
1\,+\ O(e^{-\delta/\epsilon}\,\epsilon^{\mu}).
\end{equation}
Moreover, the set $\{\chi_j(\epsilon,\,t)\}_{j=1,2}$ is orthonormal up to
errors of order $e^{-\delta/\epsilon}\,\epsilon^{\mu}$.

\noindent
{\bf 2.} The Schr\"odinger equation has solutions $\Psi_j(\epsilon,\,t)$,
$j=1,2$, such that uniformly in $t\in\mathbb R$ as
$\epsilon\rightarrow 0$,
\begin{eqnarray}\label{wow1}\nonumber
\Psi_1(\epsilon,\,t)\ =\ \chi_1(\epsilon,\,t) - {\sqrt 2}
e^{-\delta/\epsilon}\,\frac 12
\left\{
\mbox{\rm erf}\left(\frac{t}
{\sqrt{{2\,\delta\,\epsilon}}} \right)
+1\right\}\,
\chi_2(\epsilon,\,t)
\,+\, O(e^{-\delta/\epsilon}\epsilon^{\mu}),
\end{eqnarray}
and
\begin{eqnarray}\nonumber\label{wow2}
\Psi_2(\epsilon,\,t)\ =\ \chi_2(\epsilon,\,t) + {\sqrt 2}
\ e^{-\delta/\epsilon}\,\frac 12
\left\{
\mbox{\rm erf}\left(\frac{t}
{\sqrt{{2\,\delta\,\epsilon}}}\right)
+1\,\right\}\,
\chi_1(\epsilon,\,t)
\,+\, O(e^{-\delta/\epsilon}\epsilon^{\mu}).
\end{eqnarray}
\end{theorem}

\begin{remarks} 
1.\quad Recall that the function erf is defined by
\begin{equation}\label{deferf}
  \mbox{\rm erf}(x)\ =\ \frac{2}{\sqrt{\pi}}\
  \int_0^x\ e^{-y^2}\ dy\  \in\ [-1,\,1].
\end{equation}
The vectors $\chi_j(\epsilon,\,t)$, $j=1,2$, are explicitly constructed
as approximate solutions to (\ref{sch}) obtained by means of
optimal truncation of asymptotic expansions of actual solutions.
As $t\rightarrow -\infty$ they are asymptotic to the instantaneous
eigenvectors $\varphi_j(t)$ of $H(t)$, up to a phase. We call
them the optimal adiabatic states.\newline
2.\quad The transition mechanism between optimal
adiabatic states goes from the value zero to the value
${\sqrt 2}\ e^{-\delta/\epsilon}$ 
in a smooth monotonic way described by the switching function
(erf+1)/2, on a time
scale of order $\sqrt{\epsilon}$.  By contrast, the transition between
instantaneous eigenstates of the Hamiltonian displays oscillations
whose amplitudes are of order $\epsilon$ for any finite time. 
It reaches its exponentially small value only at $t=\infty$.
The vectors $\chi_j(\epsilon,\,t)$ are also optimal in that sense.
\newline
3.\quad Since the optimal adiabatic states and eigenstates
essentially coincide at $t=\pm \infty$, the transition amplitude equals
${\mathcal A}(\epsilon)\ \simeq\ {\sqrt 2}\,e^{-\delta/\epsilon}$,
up to errors of order $e^{-\delta/\epsilon}\,\epsilon^{\mu}$.
\newline
4.\quad The whole evolution operator
$U_{\epsilon}(t,\,s)$
associated with (\ref{schu}) is actually known up to errors of order
$e^{-\delta/\epsilon}\,\epsilon^{\mu}$, for any $t$ and $s$.
\end{remarks}

We prove this theorem by carefully controlling the
terms that occur in the a\-symptotic expansion of exact solutions,
so that we obtain very detailed information about the optimally
truncated expansion.
Even for our simple Hamiltonian, this involves
very delicate asymptotic analyses of
solutions to several non-linear recursion relations.

This result was soon followed by an essentially equivalent
result by Betz and Teufel \cite{BT1}, who dealt with a slightly larger
family of Hamiltonians. The main point of their analysis was
to introduce a different technique. They
defined the optimal adiabatic states as
instantaneous eigenstates of transformed Hamiltonians,
similar to (\ref{iter}),
instead of dealing with the expansions of solutions.
This way they simplified the analysis of the recursion
relations.
That paved the way for their significant generalization \cite{BT2}.
They proved that the conclusions of Theorem
\ref{supertheorem} held for general $2\times 2$ time dependent
analytic Hamiltonians. 

Let us informally state their result for real
symmetric matrices
$$
H(s)\ =\ \frac 12 
\left(\,\begin{array}{cc}Z(s)&X(s)\\X(s)&-Z(s)\end{array}\,
\right),
$$
such that $\rho(s)\ =\ \sqrt{X^2(s)+Z^2(s)}$ is strictly
positive for real $s$
and has only one zero $z_0\in\mathbb C^+$, the upper
half plane. Under certain technical assumptions,
Betz and Teufel show that the conclusion to
Theorem \ref{supertheorem} holds with $t$ given by
$$t(s)\ =\ \int_0^s\,\rho(u)\,du,$$
$\delta$ given by 
$\delta\ =\ |\mbox{Im}\ t(s)|_{s=z_0}|,$
and the prefactor $\sqrt{2}$ in front of the curly brackets
replaced by $G\neq 0$, which depends on the properties of $H(s)$ in a
neighborhood of $z_0$.

See \cite{btgiens} for a non-technical account of their findings, which
confirm Berry's intuitive arguments.


\section{Exponentially Accurate Born--Oppenheimer Approximation}

We now turn to transition phenomena in the Born-Oppenheimer
approximation.
This approximation is widely used in molecular dynamics.
It takes advantage of the disparity between the masses of nuclei
and electrons, and accordingly, of the very different time scales
for these particles' motions.
The light electrons adjust quickly to the relatively slow changes in the
configuration of the heavy nuclei. If they start in an electronic
bound state, they stay in that bound state, as if the nuclei were not
moving. This is typical adiabatic behavior. The nuclei are heavy, so
they behave in a semiclassical fashion.
The two motions are coupled because the electronic energy level
plays the role of an effective potential for the nuclei.

Hence, the Born-Oppenheimer approximation is simultaneously adiabatic
and semiclassical in nature. The breakdown of these approximations
are both associated with exponentially small phenomena: non-adiabatic
transitions on the one hand, and tunnelling on
the other hand. See {\em e.g.}, \cite{hagjoy3, hagjoy5, BR}.
It is thus reasonable
to expect the Born-Oppenheimer approximation to be valid up to
exponentially small corrections.

We set the electron masses
equal to one and take the nuclear masses proportional to
$\epsilon^{-4}$, where $\epsilon$ is a small parameter.
We let $x\in\mathbb R^d$ denote the collective positions of the nuclei
and $y\in\mathbb R^n$ denote those of the electrons. If $W(x,y)$ is the
inter-particle potential, the molecular Hamiltonian is
\begin{equation}\label{molham}
H(\epsilon)\ =\ -\,\frac{\epsilon^4}{2}\,\Delta_x\,
-\,\frac{1}{2}\,\Delta_y\,+\,W(x,y),
\end{equation}
on the Hilbert space $L^2(\mathbb R^d\times \mathbb R^n)$,
where $\Delta_z$ denotes the Laplacian in the $z$ variables. We introduce
the electronic Hilbert space $\mathcal H_{el}=L^2(\mathbb R^n)$ and the
electronic Hamiltonian $h(x)\ =\  -\,\frac 12\,\Delta_y\,+\,W(x,y)$ on 
$\mathcal H_{el}$, which depends parametrically on the position $x$ of the
nuclei. Then we can rewrite (\ref{molham}) as
\begin{equation}\label{boham}
H(\epsilon)\ =\ -\,\frac{\epsilon^4}2\,\Delta_x\,+\,h(x)
\end{equation}
In an appropriate time scale, the molecular
time-dependent Schr\"odinger equation is 
\begin{equation}\label{mol}
i\,\epsilon^2\,{\partial_t}\Psi\ =\ -\,\frac{\epsilon^4}{2}\,\Delta_x\Psi
\,+\,h(x)\,\Psi,\qquad\ \Psi \in L^2(\mathbb R^d,\mathcal H_{el} ).
\end{equation}
We study solutions to this equation in the small $\epsilon$ limit,
for times $t\in [0,T]$, where $T$ is fixed.
We never actually use the explicit form of the
Hamiltonian (\ref{molham}). We study solutions to
(\ref{mol}) with (\ref{boham}) satisfying the following two
hypotheses.
The first one says $x\mapsto h(x)$ is analytic in an appropriate sense:
\\

\noindent{\bf A$_{el}$:}\
{\em {\bf (i)} For any $x\in\mathbb R^d$, $h(x)$ is a
self-adjoint operator on some dense domain
${\mathcal D}\subset{\mathcal H}_{\mbox{\scriptsize el}}$,
where ${\mathcal H}_{\mbox{\scriptsize el}}$ is the electronic Hilbert
space. We assume the domain ${\mathcal D}$ is independent of $x$,
and that $h(x)$ is bounded below, uniformly for $x\in\mathbb R^d$.\\
\phantom{\noindent{\bf H$_0$:}\ }{\bf (ii)}
There exists $\mu>0$, such that for every
$\psi\in{\mathcal D}$, the vector $h(x)\psi$ is analytic in
$S_{\mu}\ =\ \{z\in\mathbb C^d\; :\; |\mbox{Im}(z_j)|<\mu,\;\;
j=1,\dots, d\,\}$.
}\\

The second hypothesis is the familiar gap hypothesis,
expressed this time for a space variable:\\

\noindent{\bf G$_{el}$:}
\ {\em There exists an open set $\Xi\subset\mathbb R^d$, such that for
all $x\in\Xi$, there exists an isolated,
multiplicity one eigenvalue $E(x)$ of
$h(x)$ associated with a normalized eigenvector
$\Phi(x)\in {\mathcal H}_{\mbox{\scriptsize el}}$.
}
\\

Under these two hypotheses, the Born--Oppenheimer
approximation colloquially says the following: Pick an initial
condition for (\ref{mol}) that is the product of a semiclassical
wave packet concentrated near $(a(0),\eta(0))$ in phase space,
times the electronic eigenstate $\Phi(x)$.
Then, at some later time $t>0$, the solution $\Psi$ is again given to
leading order in $\epsilon$ by a semiclassical wave packet times the
same electronic eigenstate. The nuclear wave packet is now concentrated
near the phase space point $(a(t),\eta(t))$, determined from 
$(a(0), \eta(0))$ by the classical flow generated by the classical Hamiltonian
$p^2/2+E(x)$. The spreading of the nuclear wave packet is also determined
by the classical dynamics. \\

As implied above, there is an exponentially accurate
version of the Born-Oppenheimer approximation.

\begin{theorem}\label{bigdog}\cite{hagjoy6}
Assume hypotheses {\bf A}$_{el}$ and {\bf G}$_{el}$.
There exists a vector $\Psi_*(x,t,\epsilon)\in
L^{2}({\mathbb R}^d,{\mathcal H}_{\mbox{\scriptsize el}})$
(that depends on a parameter $g$) for $t\in [0,\,T]$, such that
for small values of $g$, there exist $C(g)$
and $\Gamma(g)>0$, such that for small
$\epsilon>0$,
$$
\left\|\,e^{-itH(\epsilon)/\epsilon^2}\Psi_*(x,0,\epsilon)
\,-\,\Psi_*(x,t,\epsilon)\,
\right\|_{L^2({\mathbb R}^d,{\mathcal H}_{\mbox{\scriptsize el}})}\
\leq\ C(g)\ e^{-\Gamma(g)/\epsilon^2}.
$$
In the state $\Psi_*(x,t,\epsilon)$, the electrons have a
probability of order $\epsilon^4$ to be in the electronic subspace 
$\mbox{Span}\,\{\Phi(x)\}^\perp$.
For any $b>0$ and sufficiently small
values of $g$, the nuclei are localized near a classical path $a(t)$ in the
sense that there exist $c(g)$ and $\gamma(g)>0$, such that for small
$\epsilon>0$,
$$
\left(\,\int_{|x-a(t)|>b}\,
\|\Psi_*(x, t, \epsilon)\|^2_{{\mathcal H}_{\mbox{\scriptsize el}}}\,
dx\,\right)^{1/2}\ \leq\  c(g)\ e^{-\gamma(g)/\epsilon^2}.
$$
The nuclear configuration $a(t)$ is determined by
classical dynamics in the effective potential $E(x)$.
The limiting time $T>0$ is
determined by the condition $\{a(t),\, t\in [0, T]\}\subset \Xi$.
\end{theorem}
\begin{remarks} 
1. Under more restrictive hypotheses, the limiting time
$T$ can be allowed to grow like $|\ln(\epsilon^2)|$, the Ehrenfest
time scale. The error term becomes  $e^{-\Gamma(g)/\epsilon^\sigma}$,
where $0<\sigma<2$.\\
2. The exponentially accurate approximation has non-trivial
components in the electronic space orthogonal to the electronic
eigenvector $\Phi(x)$ which are of order $\epsilon^2$, but its
semiclassical dynamics is entirely determined by the electronic level
$E(x)$. These components of order $\epsilon^2$ correspond to the
adjustment of the spectral subspace necessary to achieve exponential
accuracy at finite times. See (\ref{sss}).
\end{remarks}

We shall not be more precise here about the construction
of the approximate solution $\Psi_*$. We do not need it,
and the details can be found in \cite{hagjoy6}.
We just mention that
$\Psi_*$ is obtained by optimal truncation of an asymptotic
series in powers of $\epsilon$ for the solution to (\ref{mol}).
Results of the same sort have then been obtained in
\cite{NenSor, MarSor, sordoni3} in more general situations and in
other contexts. Roughly speaking, these papers show that transitions
between (adjusted) spectrally isolated electronic subspaces are
exponentially small under the time evolution.
They can handle nuclear wave packets which are not
necessarily concentrated near a point of phase space in the semiclassical
limit.
In order to determine the characteristics of the nuclear component
of the solutions, a semiclassical analysis within the spectral subspaces
needs to be done. An important feature of these papers is that they
decouple the adiabatic and the semiclassical effects.

In any case, we infer from these results that transitions between
electronic subspaces are hidden in the exponentially small error term.
They are expected to stem from some
Landau--Zener mechanism. The results in the next section verify
that this expectation is correct.


\section{Non-Adiabatic Corrections to the Born--Oppenheimer Approximation}

In Theorem \ref{bigdog}, the nuclear wave packet is concentrated
near $a(t)$.
Thus, it is reasonable to expect the electrons to behave
as though their motion were governed by $h(a(t))$.
Furthermore, it is reasonable to expect the Landau--Zener
mechanism to govern their transitions, since the replacement
of $h(x)$ by $h(a(t))$ leads to a standard adiabatic approximation.


\vspace{2mm}
From the analysis \cite{hagjoy9} of a simple
Born--Oppenheimer model that we discuss below,
the main {\em qualitative} features of this argument are correct.
However, the {\em quantitative} predictions based on the use of the
Landau--Zener formula are wrong at leading order.
For example, the exponential decay rate is wrong, even if
the width of the nuclear wave packet is negligible
in the limit $\epsilon\rightarrow 0$. Some characteristics of the
energy/momentum density of the nuclear wave packet are present in
the main features of the piece of the wave function that has made
a transition.
These aspects are clearly absent from the heuristic considerations
above.

\vspace{.2cm}
The following situation is considered in
\cite{hagjoy9}: The nuclear variable $x$ belongs to $\mathbb R$ and
the electronic Hamiltonian is given by a finite order matrix
$h(x,\delta)$. We restrict attention to a
$2\times 2$ matrix for simplicity.
We assume this matrix depends analytically on $x$ and approaches
limits sufficiently rapidly as $x\rightarrow \pm\infty$.
As we saw previously, this is the most convenient
framework in which a Landau--Zener formula can be proven.
We finally assume that the
two non-degenerate electronic levels display only one avoided
crossing in a neighborhood of $(x, \delta)=(0,0)$, of the type
described by (\ref{avcr}). Here again, $\delta$ is a
supplementary parameter of the problem.  More precisely, with
the notation of {\bf A$_{el}$} for $d=1$, we assume:\\

\noindent {\bf M$_{el}$:}\ {\em
{\bf i)} For some $\delta_0>0$ and $\mu>0$, the map
$ S_\mu\times[0,\delta_0]\ni(x,\delta)\mapsto h(x,\delta)\in M_2(\mathbb C)$
is $C^2$. For any fixed $\delta$, $x\mapsto h(x,\delta)$ is analytic
in $S_\mu$.\\
\phantom{xxxx} {\bf ii)} There exist $\nu >5/2$, $c\geq 0$,
and two matrices
$h(\pm\infty,\delta)$, such that
$$
\sup_{|y|\leq\mu, \delta\in [0,\delta_0]}\ \|h(x+iy,\delta)-h(\pm\infty,
\delta)\|\
\leq\ \frac{c}{\langle x\rangle^{\nu}},
\quad \mbox{ as}\ x \rightarrow \pm\infty.
$$
The limiting matrices
$h(\pm\infty,\delta)$  are $C^2$ in
$\delta\in [0,\delta_0]$ and are non-degenerate.
}\\

For each $x\in\mathbb R$ and each $\delta\in[0,\delta_0]$,
we denote the two eigenvalues of $h(x,\delta)$
by $E_1(x,\delta)$ and $E_2(x,\delta)$. We assume:\\

\noindent {\bf AC$_{el}$:}\ {\em
{\bf i)} For $\delta>0$ and $x\in [-\infty,\,+\infty]$,
the eigenvalues satisfy $E_1(x,\delta)< E_2(x,\delta).$\\
\phantom{xxxxxx}{\bf ii)}
When  $\delta=0$, the analytic eigenvalues $E_j(x,0)$ have only
one real crossing at $x=0$.
They are labeled so that $E_1(x,0)<E_2(x,0)$ for all $x<0$.\\
\phantom{xxxxxx}{\bf iii)}
Moreover, we assume the following behavior near the avoided crossing:
$$
E_2(x,\delta)-E_1(x,\delta)\ =\ \sqrt{x^2+\delta^2+R_3(x,\delta)},
$$
where $R_3$ is a remainder of order $3$ around $(0,0)$\\
The corresponding normalized eigenvectors are denoted by
$\phi_1(x,\delta)$ and $\phi_2(x,\delta)$.
They are chosen to satisfy the phase condition
$\langle\phi_j(x,\delta),\,\partial_x\phi_j(x,\delta)\rangle
\equiv 0$.
}

\begin{remark}
In the following discussion, $\delta$ is positive and fixed,
so we often drop it from the notation. We reintroduce it
when necessary.\end{remark}

The analysis of (\ref{mol}) in this set-up starts with a
separation of variables and the study of generalized eigenvectors
$\Phi$ of $H(\epsilon)$.
These vectors depend on a real energy parameter $E$,
and they are defined by the ODE in $\mathbb C^2$:
\begin{equation}\label{genev}
\left(\,-\,\frac{\epsilon^4}{2}\,\partial_x^2\,+\,h(x)\,\right)
\,\Phi\ =\ E\,\Phi .
\end{equation}
We pick $E$ in some bounded energy interval $\Delta\subset\mathbb R$.
To ensure that scattering is possible, we assume
\begin{equation}
\inf_{E \in\Delta }\,\Delta\ >\ \sup_{x\in \mathbb R }\,E_2(x).
\end{equation}
Choosing an energy density $Q(E,\epsilon)\in\mathbb C$,
we get an exact solution to (\ref{mol}) of the form
\begin{equation}\label{sepvar}
\int_{\Delta}\,Q(E,\epsilon)\,
\Phi(x,E,\epsilon)\,e^{-itE/\epsilon^2}\,dE.
\end{equation}

From physical intuition, we anticipate
that if the energy density is sharply peaked, then when $|t|$ is large,
these solutions to (\ref{mol}) will be concentrated in
a region where $|x|$ is large.


We choose the energy density to behave like a Gaussian of width
$\epsilon^2$, centered around an initial energy
$E_0$ in the interior of $\Delta$. More precisely we assume:\\

\noindent {\bf ED:}\quad {\em
The complex valued energy density is supported on $\Delta$ and
has the form
\begin{equation}\nonumber
Q(E,\epsilon)\ =\ e^{-\,G(E)/\epsilon^2}\
e^{-\,i\,J(E)/\epsilon^2}\ P(E,\epsilon),\qquad\mbox{where} 
\end{equation}
{\bf i)} $G\in C^3(\Delta)$ is non-negative, and 
satisfies $G(E)\ =\ g\,(E-E_0)^2/2\ +\ O(E-E_0)^3$,
for some $E_0$ in the
interior of $\Delta$ and some $g>0$.\\
{\bf ii)} $J\in C^3(\Delta)$ is real-valued.\\
{\bf iii)}  $P(E,\epsilon)\in C^1(\Delta)$ satisfies
$\ \sup_{E\in\Delta,\,\epsilon \geq 0}\
\left|\,{\partial^n_E}\,P(E,\epsilon)\,\right|\,
\leq\,C_n,$\quad for $n=0,1$.\\
}

This choice implies that in the remote past or remote future
the nuclear components are
coherent states which basically propagate freely along
the different electronic levels.
Modulo the technicality of inserting a cut-off,
Gaussian coherent states have such an energy density.
For these states we use the notation of \cite{raise}:
\begin{equation}\label{gcs}
\varphi_0(A,B,\epsilon^2,a,\eta,x)\ =\
\frac 1{\pi^{1/4}\,\epsilon^{1/2}\,A^{1/2}}\
\exp\left(\,-\,\frac{B\,(x-a)^2}{2\,A\,\epsilon^2}\,+\,
i\,\frac{\eta\,(x-a)}{\epsilon^2}\,\right),
\end{equation}
where the complex numbers $A$ and $B$ satisfy the normalization
condition $\mbox{Re}\,\overline{B}A\,=\,1$.
These states are localized in position near $x=a$, and in momentum
near $p=\eta$. Their position uncertainty is $\epsilon|A|$ and their
momentum uncertainty is $\epsilon|B|$. For a thorough discussion of these
wave packets, see \cite{raise}.
Note, however, that there are states characterized by a Gaussian energy
density which are not necessarily Gaussian in space and
momentum. See Section 6 of \cite{hagjoy9}.

\vspace{2mm}
The solutions of (\ref{mol}) considered in \cite{hagjoy9} are
characterized by their behavior in the remote past. We now describe
these asymptotic states. The generalized eigenvector equation
(\ref{genev}) can be solved by means of a WKB Ansatz as follows:
Let $k_j(x,E)=\sqrt{2(E-E_j(x))}$, $j=1,2$, be the classical
momenta corresponding to the potentials $E_j(x)$. Then, the
solutions $\Phi(x,E,\epsilon)$ to the stationary equation
(\ref{genev}) can be written as
\begin{equation}\label{wkb}
\Phi(x,E,\epsilon)\ =\ \sum_{j=1,2 \atop \sigma=\pm}\,
\phi_j(x)\ \frac{c_j^\sigma(x,E,\epsilon)}{\sqrt{2 k_j(x,E)}}\
e^{-i\sigma\int_0^xk_j(y,E)dy/\epsilon^2},
\end{equation}
where 
$c_j^\sigma(x,E,\epsilon)\in\mathbb C$ are coefficients that
satisfy 
some linear ODE in $x$, for each $E\in\Delta$.
The condition {\bf M$_{el}$} ensures the existence of the limits
of $\phi_j(x)$, $k_j(x, E)$, and $c_j^\sigma(x,E,\epsilon)$,
as $x\rightarrow\pm\infty$ 
and further provides rates at which they are approached.

We assume without loss  that the ``initial''
conditions
for the coefficients $c_j^\sigma$
are fixed at $x=-\infty$ and that 
$c_j^\sigma(-\infty,E,\epsilon)$ are independent of $\epsilon$ and $E$.
It follows that the dependence in $E$ of all the above quantities is
analytic in a neighborhood of $\Delta$, and 
$c_j^\sigma(x,E,\epsilon)$ is uniformly
bounded in $(x,E,\epsilon)$.

We now introduce freely propagating states
$\psi^\sigma(x,t,\epsilon,\pm)\in L^2(\mathbb R,\,\mathbb C^2)$ that
describe the asymptotics of the solutions
$\Psi(x,t,\epsilon)$ of (\ref{mol}) as
$t\rightarrow\pm\infty$. Let
\begin{eqnarray}\label{asst}
& &\psi^\sigma(x,t,\epsilon,\pm)\ =\ \sum_{j=1,2}\,
\psi_j^\sigma(x,t,\epsilon,\pm)\\
&=&\sum_{j=1,2}\phi_j(x)\int_{\Delta}
\frac{Q(E,\epsilon)c_j^\sigma(\pm\infty,E,\epsilon)
}
{\sqrt{2k_j(\pm\infty,E)}} e^{-itE/\epsilon^2}
e^{-i\sigma(x k_j(\pm\infty,E)+
\omega_j^\pm(E))/\epsilon^2}
\,dE ,\nonumber
\end{eqnarray}
where 
$
\omega_j^\pm(E)\ = \ \lim_{x\rightarrow\infty}\int_0^{x}k_j(y,E)\, dy-xk_j(\pm\infty,E).
$
These states are linear combinations of products of free
scalar wave packets in constant scalar potentials times
eigenvectors of the electronic Hamiltonian.
Their propagation is thus governed by the various channel
Hamiltonians. With our sign convention, the states indexed by
$\sigma=-$ propagate to the right, whereas those
indexed by $\sigma=+$ travel to the left.

By construction,
$\psi^-(x,t,\epsilon,\pm)$ is likely to be a good approximation
of $\Psi(x,t,\epsilon)$ only for
$x\rightarrow \pm \infty$, {\em i.e.}, for
$t\rightarrow \pm\infty$. Similarly, $\psi^+(x,t,\epsilon,\pm)$
should be a good approximation only for $x\rightarrow \pm \infty$,
{\em i.e.} for $t\rightarrow \mp \infty$. Quantitatively, we have
\begin{proposition}\label{scat}
Assume {\bf AC$_{el}$}, {\bf M$_{el}$} and {\bf ED}.
As $t\rightarrow \pm\infty$, 
\begin{equation}\nonumber\label{prop}
\|\,\Psi(\cdot ,t,\epsilon)\,-\,
\psi^\sigma(\cdot,t,\epsilon,\mp\sigma)\,
\|_{L^2(\mathbb R^{\mp\sigma})}\ = \ O_\epsilon(1/|t|^\beta),
\end{equation}
for any $0<\beta<1/2$.
\end{proposition}

\begin{remark} 
The estimate above is valid for any value of $\epsilon>0$, and
its proof essentially relies on integration by parts.
\end{remark}

We now analyze the freely
propagating states $\psi_j^\sigma(x,t,\epsilon,\pm)$.
We do this by investigating
the small $\epsilon$ behavior of the coefficients
$c_j^\sigma(\pm\infty,E,\epsilon)$. 
We are interested in solutions associated with only
one electronic state in the remote past.
For concreteness, we choose level 2.
Proposition \ref{scat} shows that this
corresponds to the initial condition
\begin{equation}\label{inco}
c_j^\sigma(-\infty,E,\epsilon)\
=\ \delta_{j,2}\,\delta_{\sigma,-}.
\end{equation}
The component that has made the transition to level 1
in the course of the evolution,
and propagates in the positive direction
for large positive times is $\psi_1^-(x,t,\epsilon,+)$.
It is characterized by $c_1^-(+\infty,E,\epsilon)$.
We analyze solutions of the equation satisfied by $c_j^\sigma$
by means of complex WKB methods, along the lines of \cite{AJ2},
since we are in an avoided crossing regime.
At this step, we require the parameter $\delta>0$ to be
small, so that we can apply the results of \cite{AJ2}. They show
that there exist $\delta_0>0$, and
$\delta\mapsto \epsilon_0(\delta)$,
such that for $\delta<\delta_0$ and
$\epsilon<\epsilon_0(\delta)$, 
\begin{equation}
c_1^-(+\infty,E,\epsilon)\ =\
e^{-i\theta(\zeta)}\,e^{i\int_\zeta k_2(E,z)\,dz/\epsilon^2}
\left(1+O_E(\epsilon^2)\right),
\qquad \mbox{for all}\ E\in\Delta.
\end{equation}
Here $\zeta$ is a loop in the complex plane,
based at the origin, which encircles
the zero
$z_0(\delta)\in\mathbb C\setminus \mathbb R$
of $k_1(z,E)-k_2(z,E)$ closest to the real axis.
Also, $\theta(\zeta)\in \mathbb C$ only depends on the
analytic continuation along $\zeta$ of the electronic
eigenvectors.

We now have everything to state the main result
of \cite{hagjoy9}:
\begin{theorem}
\label{mai}
Let $\psi(x,t,\epsilon)$ be a solution to the
molecular Schr\"odinger equation (\ref{mol})
with electronic Hamiltonian $h(x,\delta)$
satisfying hypotheses {\bf M$_{el}$} and {\bf AC$_{el}$}.
Assume the solution is characterized
asymptotically in the past for negative $x$'s by
$$
\lim_{t\rightarrow -\infty}\
\|\,\psi(x,t,\epsilon)\,-\,\psi^-_2(x,t,\epsilon,-)
\,\|_{L^2(R^-)} \ =  0,
$$
with
$$
\psi^-_2(x,t,\epsilon,-)\ =\ \phi_j(x)\
\int_{\Delta}\ \frac{Q(E,\epsilon)}{\sqrt{2k_2(-\infty,E)}}\
e^{-itE/\epsilon^2}\,
e^{i(xk_2(-\infty,E)+\omega_2^-(E))/\epsilon^2}\,dE,
$$
where the energy density $Q(E,\epsilon)$ satisfies {\bf ED}.
Let $E(k)=k^2/2+E_1(+\infty)$ 
\begin{eqnarray}\label{al}\nonumber
\alpha(E)&=&G(E)\,+\, \mbox{Im}\, \smallint_{\zeta}\,k_2(z,E)\,dz,
\qquad \mbox{and}\\
\label{ka}\nonumber
\kappa(E)&=&J(E)\,-\, \mbox{Re}\,\smallint_{\zeta}\,k_2(z,E)\,dz\,-\,
\omega_1^+(E).
\end{eqnarray}
Assume $E^*$ is the unique absolute minimum of $\alpha(\cdot)$
and set $k^*=\sqrt{2(E^*-E_1(\infty))}$.

Then, there exist $\delta_0>0$, $p>0$
arbitrarily close to 5/2, and a function
$\epsilon_0:(0,\delta_0)\rightarrow \mathbb R^+$,
such that for all $0<\beta<1/2$,
$\delta <\delta_0$, and $\epsilon<\epsilon_0(\delta)$,
the following asymptotics hold as $t\rightarrow \infty$:
\begin{eqnarray}\nonumber
&&\hspace{-8mm}\psi_1^-(x,t,\epsilon, +)\ =\\ \nonumber
&&\quad \phi_1(x)\ 
\frac{e^{-i\theta(\zeta)}
\epsilon^{3/2}\pi^{3/4}
e^{-\alpha(E^*)/\epsilon^2}}{\left(\frac{d^2}
{dk^2}\alpha(E(k))|_{k^*}\right)^{1/4}}\ \,
e^{iS_+(t)/\epsilon^2}\ 
\varphi_0(A_+(t),B_+,\epsilon^2,a_+(t),\eta_+,x)
\\
&&\quad \times\
P(E^*,\epsilon)\ \sqrt{k^*}\ 
e^{-i(\kappa(E^*)-{k^*}^2\kappa'(E^*))/\epsilon^2}\ +\
O(e^{-\alpha(E^*)/\epsilon^2}\epsilon^{p}) 
\,+\,O_\epsilon
\left(1/t^\beta\right),\nonumber
\end{eqnarray}
where $\varphi_0$ is a Gaussian (\ref{gcs}) parametrized by
\begin{eqnarray}\nonumber
&&\eta_+=k^*,\qquad
a_+(t)=\kappa'(E^*)+k^*t,\qquad
B_+=1/\sqrt{\partial_k^2 \alpha(E(k))|_{k=k^*}}\ ,
\\ \nonumber\label{ident}
&&A_+(t)=B_+\,\left(\,\partial_k^2\alpha(E(k))|_{k=k^*}+i(
\partial_k^2\kappa(E(k))|_{k=k^*}+t)\,\right),
\qquad\mbox{and}\\ \nonumber
&&S_+(t)=\left(\,{k^*}^2/2 - E_1(+\infty)\,\right)\ t.
\end{eqnarray}
All error terms are estimated in the $L^2(\mathbb R)$ norm, and
the estimate $O(e^{-\alpha(E^*)/\epsilon^2}\epsilon^{p})$
is uniform in $t$. The $O_\epsilon(1/t^\beta)$
may depend on $\epsilon$.
\end{theorem}

\begin{remark}The leading order of the piece transmitted
to level 1 is always
given by an exponentially small prefactor times
a freely propagating Gaussian nuclear wave packet,
{\it i.e.} a Gaussian semiclassical wave packet in
the constant potential given by the asymptotic electronic level.
This is true even if the incoming state is not Gaussian.
\end{remark}

All quantities computed from the electronic
Hamiltonian depend on $\delta$,
despite this not being emphasized in the notation. 
Although we shall not do it here,
one can further investigate the characteristics 
of the asymptotic state for small
$\delta$'s. Without being too precise,
we mention that if $k_c(E)=\sqrt{2(E-E_2(0,0))}$
is the classical momentum at
the crossing point of the electronic levels for $\delta=0$,
one has
$$\alpha(E)=g(E-E_0)^2/2+
\frac{\delta^2\pi}{4\,k_c(E)}+O((E-E_0)^3)+O(\delta^3).$$
The factor $\delta^2\pi/(4\, k_c(E))$ is the Landau--Zener decay rate
computed from the electronic levels divided by the classical momentum
at the avoided-crossing,
and $g>0$ stems from the energy density.
So, for small values of $\delta$ and for a sufficiently
narrow energy window $\Delta$,
$\alpha$ is a quadratic term plus a positive,
decreasing, convex function of $E$.
This implies that $E^*$, determined
by $\alpha'(E^*)=0$, is strictly larger
than $E_0$ and depends explicitly on the energy density.
Furthermore, the decay rate has
$\alpha(E^*)<\alpha(E_0)$. This has the following consequences.
\begin{remark}
The average momentum $k^*$ is larger than what a
na\"\i ve application of energy conservation
yields, and the exponentially small prefactor
is larger than the prediction
given by the Landau--Zener formula for $h(a(t))$.
The correct values of both of these quantities
depend explicitly on the chosen energy density.
\end{remark}

The lesson we learn from this is that the leading
behavior of the piece of the wave function that describes
transitions from one electronic level to the other cannot be
determined by straightforward semiclassical considerations and
the Landau--Zener formula. The
way the incoming asymptotic state is prepared
in the remote past plays a crucial role in the main
characteristics of the asymptotics.

Results of this sort are obtained in \cite{joymar}
for scattering systems defined by more general
semiclassical autonomous PDEs entertaining dispersive waves.
While the overall general strategy
remains the same, the separation of variables,
study of the stationary equation
by means of complex WKB methods and
analysis of the space-time properties
of exact solutions, can be quite different.

The description of the time development of the transitions
between the electronic levels represents a challenging question.
Betz and Teufel \cite{bt3} have heuristic arguments and
numerical evidence supporting a description of this phenomenon.
However, the mathematical problem remains open.

\end{document}